\documentclass[a4paper,reqno]{amsart}

\usepackage{amsthm}\usepackage{amsaddr}
\usepackage{amsmath}\usepackage{amssymb}

\usepackage{a4wide}

\usepackage[backref]{hyperref}

\theoremstyle{definition}

\renewcommand{\d}{{\mathrm{d}}}

\newcommand{\J}{{\mathbb J}}
\newcommand{\1}{{\mathbb I}}
\newcommand{\diag}{\mathrm {diag}}
\newcommand{\ab}{{a}}
\newcommand{\bb}{{b}}
\newcommand{\xx}{{\mathbf x}}
\newcommand{\XX}{{\mathbf X}}
\newcommand{\PP}{{P}}
\newcommand{\KK}{{K}}
\newcommand{\pp}{{\mathbf p}}
\newcommand{\kk}{{\mathbf k}}

\newcommand{\medbox}[1]{\fbox{%
    \rule[-10pt]{0pt}{25pt}$\;\;\displaystyle{#1}\;\;$}%
}

\newcommand{\eps}{\varepsilon}

\title[Memory effect in impulsive waves: comments, corrections, 
clarifications]{The memory effect in impulsive plane waves:\\ 
comments, corrections, clarifications${}^0$}
\author{Roland Steinbauer}
\address{University of Vienna, Faculty of Mathematics\\Oskar-Morgenstern-Platz 
1, A-1090 Wien, Austria}
\email{roland.steinbauer@univie.ac.at}

\begin{document}

\begin{abstract}
Recently the ``memory effect'' has been studied in plane gravitational waves 
and, in particular, in \emph{impulsive} plane waves. Based on an analysis of 
the particle motion (mainly in Baldwin-Jeffery-Rosen coordinates) a ``velocity 
memory effect'' is claimed to be found in \cite{ZDH:impulsive}. 
Here we point out a conceptual mistake in this account and employ earlier 
works to explain how to correctly derive the particle motion and how to 
correctly deal with the notorious distributional Brinkmann form of the metric 
and its relation to the continuous Rosen form. 
\vskip 1em
  
\medskip
    
\end{abstract}

\maketitle

\footnotetext[0]{This note was published in the present form in CQG as a 
`comment' with the title changed to  {\em Comment on 'Memory effect for 
impulsive gravitational waves'}. The precise reference is Class.\ Quant.\ Grav.\ 
{\bf 36}(9) 098001, 14, (2019).}

\section{Introduction}
 The ``wave memory effect'', see e.g.\ \cite{ZP,BG,BT,C} has recently attracted 
much interest  due to its possible experimental detection and a growing 
number of publications addresses this topic, see e.g.\ the intoduction of 
\cite{ZDH:impulsive} and the literature cited therein. In a recent series of 
papers \cite{ZDGH:pl,ZDGH:soft,ZEGH:18} the ``memory effect'' has been studied 
in plane gravitational waves and in \cite{ZDH:impulsive} these studies have 
been extended to \emph{impulsive} plane waves. 

Impulsive plane waves and, more generally, impulsive pp-waves have been 
introduced by Roger Penrose in the late 1960-ies, see \cite{P:68,P:68a}, and 
\cite{P:72} for a more extensive treatment. They are spacetimes of low 
regularity, described alternatively by a (locally Lipschitz) continuous metric 
in (Baldwin-Jeffery-)\footnote{On historical grounds the name 
``Baldwin-Jeffery-Rosen coordinates'' seems only to be accurate in the context 
of plane waves.} Rosen form, or by a distributional metric in Brinkmann form. 
Over the years they have attracted the attention of researchers in exact 
spacetimes (who have widely generalized the original class of solutions), of 
mathematicians (who used them as relevant key-models in low reguarity Lorentzian 
geometry), and of particle physicists (who have considered quantum scattering in 
these geometries). 

In their work ``Memory effect for impulsive gravitational waves'' the authors of 
\cite{ZDH:impulsive} derive the geodesics in impulsive plane waves using both 
forms of the metric. They find that in both coordinate systems particles 
initially at rest suffer a jump in their transversal velocities when crossing 
the impulse and start to move apart with constant speed. From this they conclude 
the occurrence of a ``velocity memory effect''. While this behavior of the 
geodesics \emph{in Brinkmann coordinates} is certainly correct and in accordance 
with the literature, the corresponding claim for the geodesics \emph{in Rosen 
coordinates is incorrect. }In this note we analyse in detail the approach of 
\cite{ZDH:impulsive} to explain why this derivation of the geodesics in Rosen 
coordinates is flawed and leads to inconsistent results. In particular, we 
address the conceptual intricacies that originate from the low regularity nature 
of the spacetimes involved. Moreover---based on earlier works---we show how 
to calculate the geodesics in Rosen coordinates and how to use their 
$C^1$-regularity to employ a ``$C^1$-matching procedure'' that leads to 
transparent ``jump formulas'' in Brinkmann coordinates. Finally, we explain how 
one can handle the subtle interrelations between the two forms of the metric in 
a mathematically meaningful way.
\medskip

Throughout this note we have strived for maximum clarity at the expense of 
brevity. We aim at completely resolving the situation and we express our hope 
that in this way we may prevent further confusion in the literature\footnote{As 
of Mar.\ 14, 2019, Inspire counts already 15 citations for the paper 
\cite{ZDH:impulsive}.}. 
\bigskip

In the remainder of this section we introduce our notation.
Generally we follow the notations and conventions of \cite{ZDH:impulsive} to 
make comparisons simple.   
The \emph{metric} of plane gravitational waves in Brinkmann
coordinates\footnote{Abbreviated as (B)-coordinates from now on.} 
$X^\mu=(\XX,U,V)$ with $\XX=(X^i)=(X,Y)$ is written as\footnote{Again,  
to simplify comparison, we have taken equation numbers 
to coincide with \cite{ZDH:impulsive}.} 
\begin{equation}\label{eq:generalB}\tag{2.1}
 \d s^2=\delta_{ij}\d X^i\d X^j+2\d U\d V+K_{ij}(U)X^iX^j\d U^2,
\end{equation}
with the profile fixed to the $+$ polarisation, hence given by 
\begin{equation}\label{eq:profileB}\tag{2.2}
 {K}=(K_{ij})=\frac{1}{2}\,\mathcal{A}(U)\, 
\diag(1,-1)=\frac{1}{2}\mathcal{A}(U)\,\mathbb{J},
\end{equation}
where the dependence of the function ${\mathcal A}$ on retarded time is 
arbitrary but smooth and we use the abbreviation $\mathbb{J}=\diag(1,-1)$. 
On the other hand, in Baldwin-Jeffery-Rosen coordinates\footnote{Abbreviated as 
(R)-coordinates below.} $x^\mu=(\xx,u,v)$ with $\xx=(x^i)=(x,y)$ the metric is 
written as
\begin{equation}\label{eq:generalR}\tag{2.4}
 \d s^2=a_{ij}(u)\d x^i\d x^j+2\d u\d v.
\end{equation}
Here the profile ${a}=(a_{ij})$ is a positive definite $(2\times 
2)$-matrix which depends again arbitrary but smoothly on retarded time $u$.

The \emph{transformation} between the (R) and (B)-coordinates is written as
\begin{equation}\label{eq:generalTRSF}\tag{2.6}
 \XX=\PP(u)\xx,\quad U=u,\qquad V=v-\frac{1}{4}\xx\cdot\dot\ab(u)\xx,
\end{equation}
where the $(2\times 2)$-matrix $\PP(u)$ is a square root of $\ab$, i.e.,
\begin{equation}\label{eq:aP}\tag{2.5}
 \ab(u)=\PP(u)^T\PP(u)
\end{equation}
when going from (R) to (B)-coordinates, and a solution of the Sturm-Liouville problem
\begin{equation}\tag{2.7}
 \ddot\PP=\KK\PP,\qquad \PP^T\dot\PP-\dot\PP^TP=0,
\end{equation}
if one goes from (B) to (R)-coordinates. The profiles are related via
\begin{equation}\tag{2.8}
 \KK=\frac{1}{2}\PP\Big(\dot\bb+\frac{1}{2}\bb^2\Big),\quad
 \mbox{with}\ \bb=\ab^{-1}\dot\ab,
\end{equation}
and so the Ricci-flat condition becomes
$
 \mathrm{tr}\,\KK=0\ \Leftrightarrow\ \mathrm{tr}\,(\dot\bb+\frac{1}{2}\bb^2)=0.
$
\medskip

The respective \emph{impulsive limits} are written with profiles 
\begin{align}\tag{2.16}\label{eq:ip1}
 {a}(u)&={P}^2(u)=(\1+u_+c_0)^2=(\1+u_+k\J)^2,\\
 \mathcal{A}(u)&=2k\delta,\tag{2.17}\label{eq:ip2}
\end{align}
where $k$ is the positive eigenvalue of the symmetric $(2\times 2)$-matrix 
$c_0$, which arises by solving for the flatness condition in the after-zone of 
the wave, cf.\ \cite[p.\ 4]{ZDH:impulsive}.  Sometimes we will set $k$ to 
$k=1/2$ or to $k=1$ and as usual we have put $u_+=0$, $(u\leq 0)$ and $u_+=u$ 
($u\geq 0)$. Finally, $\delta$ is the Dirac-measure.

\section{The peculiarities of impulsive wave spacetimes}\label{sec:interlude}
Here we pause for a moment and recall an essential issue in the construction of 
impulsive wave spacetimes, for an extensive review see \cite[Ch.\ 20]{GP:09} and 
\cite{P:02} or the more (astro-)physically oriented monograph \cite{BH:03}. 
These spacetimes were introduced by Roger Penrose in \cite[p.\ 189ff]{P:68},  
\cite[p.\ 82ff]{P:68a} using what he called a ``scissors and paste'' method: two 
Minkowski half-spaces $M^\pm$ are glued along a null hyperplane with a ``warp'', 
i.e., a shift along the generators of the hyperplane when one passes from 
$M^-$ to $M^+$. Despite the fact that the Brinkmann form of the impulsive 
pp-wave contains a distributional component, the spacetime is actually  (locally 
Lipschitz\footnote{The Lipschitz property is decisive, since it guarantees the 
connection to be locally bounded. Moreover, it has recently been noted that 
Lipschitz-regularity of the metric is a threshold, below which even the most 
basic facts of causality theory fail to hold \cite{CG:12,GKSS:19}.}) continuous 
but not $C^1$. This is seen from the continuous Rosen form of the metric first 
given in (the plane wave case in) \cite[p.\ 103]{P:72} which possesses a locally 
bounded connection and a distributional curvature component $\Psi_4$.
 
\medskip

Let us emphasise the fact that this construction does \emph{not provide a 
global background Minkowski space} on which the wave impulse can be thought to 
travel. This fact is somewhat obscured by the use of the so-called 
\emph{Souriau-coordinates} \cite{Sou:73}, $\hat{\xx}^\mu=(\hat{\xx},\hat u,\hat 
v)$ with 
$\hat{\xx} =(\hat x^i)=(\hat x,\hat y)$ which are defined to be such that the 
metric is manifestly Minkowskian in both halves. Consequently, in the impulsive 
case, the transformation between (S) and (R)-coordinates is the identity for 
$u=\hat u<0$, and for $u=\hat u>0$ it is given by (cf.\ \eqref{eq:generalTRSF})
\begin{equation}\tag{2.25}\label{eq:2.25}
 \hat\xx=\PP(u)\xx,\quad \hat u=u,\quad \hat 
v=v-\frac{1}{4}\xx\cdot\dot\ab(u)\xx=v-\frac{1}{2}\xx\cdot 
c_0\PP(u)\xx=v-\frac{1}{2}\xx\cdot k\J\PP(u)\xx,
\end{equation}
where we have used that in the impulsive case we have 
$\dot\ab=2\PP\dot\PP=2c_0\PP$ (cf.\ \eqref{eq:aP}, \eqref{eq:ip1}). 
\medskip

A useful way of thinking about the (S)-coordinates and the whole construction 
of impulsive spacetimes is the line of argument given in  \cite[Sec.\ 
20.2]{GP:09}, for the present setting cf.\ \cite[eq.\ (9)]{LSS:14}: 
Starting from Minkowski space with global coordinates now denoted 
by\footnote{The coordinates $\tilde x^\mu$ correspond to 
  $(x,y,{\mathcal U},{\mathcal V})$ in \cite{LSS:14} with a change in 
sign in $\tilde v$ w.r.t.\ $\mathcal{V}$, which is due to the choice of 
$-2\d\mathcal{U}d\mathcal{V}$ in the metric \cite[eq.\ (1)]{LSS:14}.
Moreover, the (R)-coordinates $x^\mu$ correspond to $(X,Y,U,V)$ in \cite{LSS:14}.} $\tilde 
x^\mu=(\tilde\xx,\tilde u,\tilde v)$ with $\tilde\xx=\tilde x^i=(\tilde 
x,\tilde 
y)$, i.e\footnote{Equations \emph{not} appearing in \cite{ZDH:impulsive} are 
numbered consecutively.}.,
\begin{equation}\label{eq:manifest_minkowski}
 \d s^2=\d\tilde\xx^2+2\d\tilde u\d\tilde v
\end{equation}
one uses the identity for $\tilde u<0$ and the transformation
\begin{align}\label{eq:trsfsplit}\nonumber
 \tilde u&=u,\\\nonumber
 \tilde \xx&=\xx+u\partial H=(\1+u_+k\J)\xx=\PP(u)\xx,\\ 
 \tilde v&=v-H-\frac{1}{2}u\partial H\cdot\partial H
   =v-\frac{k}{2}\xx\cdot\xx-\frac{1}{2}uk\xx\cdot\xx
   =v-\frac{1}{2}\xx\cdot k\J(\1+uk\J)\xx\\\nonumber
  &=v-\frac{k}{2}\xx\cdot \J\PP(u)\xx
\end{align}
for positive $\tilde u$. Here we have set the profile $H$ in \cite[eq.\ (9)]{LSS:14} to $H=(k/2)(x^2-y^2)=(k/2)\xx\cdot\J\xx$ so that $\partial H=k\J\xx$ and $\partial H\cdot\partial H=k^2\xx\cdot\xx$ and we see that we obtain exactly \eqref{eq:2.25}.

However, let us now combine the transformation \eqref{eq:trsfsplit} for $\tilde 
u>0$ with the identity for $\tilde u<0$ formally to the \emph{discontinuous} 
transformation valid \emph{for all values of} $\tilde u$ resp.\ $U$
(which was also first given by Penrose explicitly in \cite{P:72})
\begin{align}\label{eq:trsfdiscont}
 \tilde u&=U,\nonumber\\
 \tilde \xx&=\xx+U_+\partial H=\PP(U)\xx,\\\nonumber
 \tilde v&=V-\theta(U)H-\frac{1}{2}U_+\partial H\cdot\partial H
   =V-\frac{k}{2}\theta(U)\J\xx\cdot(\1+U_+k\J)\xx
   =V-\frac{k}{2}\theta(U)\xx\cdot \J\PP(U)\xx.
\end{align}
Here we have used the identity of $L^\infty$-functions $\theta(U)U_+=U_+$.
\medskip

Now if one \emph{formally} transforms the metric \eqref{eq:generalR} with the 
impulsive profile \eqref{eq:ip1} according to \eqref{eq:trsfdiscont} 
\emph{keeping the distributional terms} then one obtains precisely the (B)-form 
\eqref{eq:generalB} of the metric with the distributional profile 
\eqref{eq:ip2}. Here we say ``formally'' since in addition to the standard 
distributional identities
\begin{equation}\label{eq:di}
 (U_+)'=\theta, \quad\mbox{and}\quad \theta'=\delta 
\end{equation}
one has to use the ``multiplication rules'' 
\begin{equation}\label{eq:mr}
 \theta u_+=u_+,\quad \theta^2=\theta,\quad\mbox{and}\quad \theta\delta=\frac{1}{2}\delta
\end{equation}
which come from the grey areas of distribution theory: a careless combination of \eqref{eq:di} and \eqref{eq:mr} easily leads to contradictions as in
\begin{align}
 \theta^2=\theta\ \Rightarrow 2\theta\theta'=2\theta\delta=\delta\
 \Rightarrow \theta\delta=\frac{1}{2}\delta,\quad \mbox{but}\quad 
 \theta^3=\theta \ \Rightarrow 3\theta^2\theta'=3\theta\delta=\delta\
 \Rightarrow \theta\delta=\frac{1}{3}\delta.
\end{align}
Structurally speaking the problems arise when one combines rules like 
$\theta^2=\theta$, which perfectly hold for $L^\infty$-functions with taking 
derivatives---which then has to be carried out in the sense of distributions.

The above procedure, however, has been made mathematically rigorous 
(using nonlinear distributional geometry in the sense of the geometric theory 
of generalized functions \cite{GKOS:01}) even in the \emph{pp}-wave case in 
\cite{KS:99}, see also Section \ref{sec:3.7}, below.
\medskip

We summarize with the following \emph{observation and warning}: The 
(S)-system does not cover the whole manifold given by the (B)-coordinates (or 
the (R)-coordinates near the null surface $\{U=0\}$) but is only valid for all 
$U=u=\hat u\not=0$ and actually consists of two patches which do not overlap and 
can only be joined via the (B)- or (R)-coordinates.

\section{Particle motion}
We now turn to the heart of the matter, i.e., the geodesics in impulsive 
plane waves. We will intensively comment on the approach 
taken in \cite[Section 5]{ZDH:impulsive}. Since in this approach the symmetries 
of the spacetime are used in an essential way, we start with a general comment 
on this strategy.

\subsection{A general remark on the use of symmetries}
The symmetries of \emph{extended} (i.e., non-impulsive) plane waves are 
identified in \cite{DGHZ:caroll} to be a subgroup of the Carroll group. 
Using the symmetries, the geodesics (again in the 
\emph{extended} case) can be calculated efficiently, especially in 
(R)-coordinates. This has been done in \cite[Sec.\ 3.2]{DGHZ:caroll}, 
see also \cite[eqs.\ (2.11)]{ZDGH:pl} and \cite[Sec.\ IV B]{ZDGH:soft}. The 
constants of motion are given by
\begin{equation}
 \mathbf{p}=a(u)\dot\xx,\quad \mathbf{k}=\xx(u)-H(u)\mathbf{p},
\end{equation}
where already the 5th conserved quantity $\mu=\dot u$ was used to parametrise 
the geodesics by the coordinate $u$. Here the matrix-valued function $H$ is 
given by\footnote{To be consistent with \cite{ZDH:impulsive} we here keep the 
letter $H$---this is, however, \emph{not} to be confused with the $H$ of Section 
\ref{sec:interlude}.} 
\begin{equation}\tag{4.2}\label{eq:4.2}
 H(u)=\int\limits_0^ua^{-1}(w)\,dw.
\end{equation}
Additionally, the kinetic energy
\begin{equation}
 e=\frac{1}{2}g_{\mu\nu}\dot x^\mu\dot x^\nu
 =\frac{1}{2}\dot\xx\cdot a(u)\dot\xx+\dot v
 =\frac{1}{2}\mathbf{p}\cdot a^{-1}(u)\mathbf{p}+\dot v
\end{equation}
is conserved and is set to $e=-1,0,1$ in the timelike, null and spacelike 
case respectively. This finally leads to the explicit expression for the 
geodesics, cf.\ \cite[eqs.\ (3.11)]{DGHZ:caroll}, \cite[eqs.\ (2.11)]{ZDGH:pl}
\begin{equation}
  \xx(u)=H(u)\pp+\kk,\quad v(u)=-\frac{1}{2}\pp\cdot H(u)\pp+eu+d,
\end{equation}
where $d$ is a constant of integration.
\medskip

However, while in the extended case the symmetries allow one to nicely express 
the geodesics, it seems less obvious that this is also an efficient approach in 
the impulsive case. There the particle motion off the impulsive hypersurface 
is trivial and if one wants to derive its form in (R)-coordinates this can be 
done using the transformation \eqref{eq:generalTRSF} or by several other simple 
means, see Section \ref{sec:3.5}. This raises the question:
\begin{enumerate}
\item[(Q1)] Why use symmetries to calculate the geodesics in the 
\emph{impulsive} case at all?
\end{enumerate}
As a follow-up, a more subtle question arises from the further procedure 
applied in \cite[Sec.\ 5.1]{ZDH:impulsive} (discussed in the next section)
\begin{enumerate}
\item[(Q2)] How to relate the values of the constants of motion in the before- and after-zones of the impulsive wave and how to argue for that?
\end{enumerate}
We will come back to this questions later on in Section \ref{sec:3.5} but first have a 
closer look on the just mentioned procedure.

\subsection{The approach of section 5.1 in \cite{ZDH:impulsive}} 
In the main section 5 of \cite{ZDH:impulsive} the geodesics are computed. In 
particular, in section 5.1 this is done in (R)-coordinates with a 
substantial use of (S)-coordinates. To start with, the form of the geodesics is 
derived using the symmetries, where the constants of motion are allowed to be 
different in the before- and the after-zones of the wave, denoted by $\pp_\pm$, 
$\kk_\pm$, and $e_\pm$, cf.\ \cite[eqs.\ (4.12), (4.13)]{ZDH:impulsive}. This 
readily leads to the form of the geodesics in the before- and the after-zones 
(again using the $\pm$-notation) 
\begin{align}\tag{5.2}\label{eq:5.2}
 \xx_\pm(u)=\kk_\pm+H(u)\pp_\pm,\quad
 v_\pm=-\frac{1}{2}\pp_\pm\cdot H(u)\pp_\pm+e_\pm u+d_\pm,
\end{align}
where now $d_\pm$ are constants of integration.

Then on physical grounds the geodesics are assumed to be continuous across $u=0$ such that
\begin{equation}\tag{5.3}\label{eq:5.3}
 \kk_\pm=\xx(0)=:\xx_0,\quad d_\pm=v_\pm(0)=:v_0,
\end{equation}
where the choice that $H(0)=0$, cf.\ equation \eqref{eq:4.2} was used as well 
as the fact that $a(0)=1$, cf. equation \eqref{eq:ip1}. Again using the latter 
equation one moreover obtains from \eqref{eq:5.2} that 
\begin{equation}\tag{5.4}\label{eq:5.4}
 \pp_\pm=\dot\xx_\pm(0),\quad \dot v_\pm(0)=e_\pm-\frac{1}{2}|\pp_\pm|^2,
\end{equation}
where these remaining constants of motion are still allowed to be different in 
the two zones. Finally one may explicitly calculate $H$ from equations 
\eqref{eq:4.2} and \eqref{eq:ip1} to obtain 
\begin{equation}\tag{4.4}\label{eq:4.4}
 H(u)=u_-\1 + u_+\PP^{-1}(u).
\end{equation}
This gives the following form of the geodesics
\begin{align}
 \xx(u)&=\xx_0+u_-\dot\xx_-(0)+u_+\PP^{-1}(u)\dot\xx_+(0),
 \nonumber\\\tag{5.7}\label{eq:5.7}
 v(u)&= v_0+u_-\dot v_-(0)+u_+\dot v_+(0)
       +\frac{1}{2}\dot \xx_+(0)\cdot u_+(\1-\PP^{-1}(u))\dot\xx_+(0).
\end{align}
These geodesics are determined by the following set of $9$ real
constants $\xx_0$, $v_0$, $\dot\xx_\pm(0)$, $\dot v_\pm(0)$, while 
there should only be $6$ since we have to subtract the two constants $u(0)=0$ 
and $\dot u(0)=1$, which we have used to write the geodesics in the above form, from the usual $8$ initial positions and speeds in a $4$-dimensional spacetime.

To determine at least some of the ``spurious constants'' the authors of 
\cite{ZDH:impulsive} now employ the (S)-coordinates and limit their 
considerations to the case of particles being at rest in the before-zone, a 
condition that can be expressed in the (S)-system of coordinates to be
\begin{equation}\tag{5.9}\label{eq:5.9}
 \hat\xx(\hat u)=\hat\xx_0,\quad
 \hat v(\hat u)=\hat v_0+e\hat u.
\end{equation}
Their decisive argument now is (cf.\ \cite[p.\ 10, bottom]{ZDH:impulsive}):
\begin{align}\nonumber
  &\mbox{\it ``Now, the after-zone being also flat and indeed Minkowskian when 
    the     hatted}\\[-.5em] \tag{*}\label{eq:argument}
&\mbox{\it (S) coordinates are used, we argue that the latter have the same 
  parametric}\\[-.5em]
  &\mbox{\it form in the after-zone: (5.9) holds for all u.''}\nonumber
\end{align}
Using this argument the authors of \cite{ZDH:impulsive} go on to rewrite the 
geodesics in (R)-form. To this end they apply the inverse of the transformation 
\eqref{eq:2.25} for $\hat u=u>0$ with the impulsive profile \eqref{eq:ip1} inserted explicitly, i.e.,
\begin{equation}\tag{5.8}\label{eq:5.8}
  \xx=(\1+uk\J)^{-1}\hat\xx,\quad
  \hat u=u,\quad
  v=\hat v+\frac{1}{2}\hat\xx\cdot k\J(\1+uk\J)^{-1}\hat\xx,
\end{equation}
which implies the relations of the constants (just set $u=0$)
\begin{equation}
  \hat\xx_0=\xx_0,\quad\hat v_0=v_0-\frac{1}{2}\xx_0\cdot k\J\xx_0
\end{equation}
and obtain the following geodesics
\begin{align}
  \xx(u)&=
  (\1+u_+k\J)^{-1}\xx_0,\nonumber\\
  \tag{5.10}\label{eq:5.10}
  v(u)&
  = v_0+eu-\frac{1}{2}\xx_0\cdot k\J\Big(\1-(\1+u_+k\J)^{-1}\Big)\xx_0.
\end{align}
 
\subsection{Interpretation of the results and comments}\label{sec:3.3}

After deriving the above result \eqref{eq:5.10} the authors of 
\cite{ZDH:impulsive} say: 

{\it 
``We see that the geodesic equation (5.10) is a 
special case of (5.7) where the after-zone initial velocity has been fixed by 
the initial conditions $\xx(0) =\xx_0$ and $\dot\xx(0-) [=\dot\xx^-_0]= 0$, namely
\begin{equation}\tag{5.11}
\medbox{\ \dot\xx(0+)[=\dot\xx^+_0]=-c_0\xx_0.\ }
\label{5.11}
\end{equation}
The impulsive GW induces a (sort of)  \textit{``percussion''}
\cite{Sou:73}, since
  \begin{align}
    \label{eq:5.12a}\tag{5.12a}
    \bigtriangleup\dot\xx&=\dot\xx(0+)-\dot\xx(0-)=-c_0\xx_0,
    \\\tag{5.12b}
    \label{eq:5.12b}
   \bigtriangleup\dot{v}&=\dot{v}(0+)-\dot{v}(0-)=-\frac{1}{2}\vert{}
   c_0\xx_0\vert^2.\,\mbox{''}
  \end{align}
  \label{percussion}
}

First we note that the geodesics \eqref{eq:5.10} are continuous not
because {\it ``[...] this follows from (5.10).''}, see \cite[Sec.\ 5.1, last 
paragraph]{ZDH:impulsive} but because this was assumed during the procedure 
explicitly in \cite[p.\ 10, first lines]{ZDH:impulsive} and mentioned here prior to \eqref{eq:5.3}.
\medskip

The main issue is, however, that the geodesics \eqref{eq:5.10} are \emph{not} 
continuously differentiable and, moreover, that they are actually \emph{not the 
correct geodesics in the first place}, as we are going to argue in the 
following:

First, it was shown in \cite[Thm.\ 1]{LSS:14} that the geodesics even in all 
impulsive (non-plane) \emph{pp}-waves are $C^1$-curves. There, Carath\'eodory's 
solution concept is used, which is the most natural extension of classical 
ODE-theory using Lebesgue theory of integration. In fact, instead of solving the 
ODE with an $L^\infty$-right hand side one solves the classically equivalent 
integral equation, see e.g. \cite[Ch.\ 3 \S 10, Suppl.\ 2]{W:98}. This is 
actually a special case of the fact that the geodesics\footnote{In the sense of 
Filippov \cite{Fil:88},  which is the appropriate solution concept there.} of 
any locally Lipschitz continuous metric are $C^1$-curves, see \cite{Ste:14}. 
\medskip 

Second\footnote{If one remains sceptical about the use of appropriate solution 
concepts for ODEs with $L^\infty$-right hand side, like Carath\'eodory's, the 
following argument might be even more convincing.}, we explicitly
demonstrate that the geodesics of \eqref{eq:5.10} \emph{do not} 
satisfy the geodesic equation. Let us demonstrate this just for the 
$x^1=x$-component, which reads 
\begin{equation}\label{eq:10}
  x(u)=\frac{x_0}{1+ku_+}.
\end{equation}
We find (using \eqref{eq:mr}, as well as formally applying the product rule) 
\begin{equation}
  \dot x(u)=-\frac{kx_0\theta(u)}{(1+ku_+)^2},\quad
  \ddot x(u)=-\frac{kx_0\delta(u)}{(1+ku_+)^2}+\frac{2k^2x_0\theta(u)}{(1+ku_+)^3}.
\end{equation}
Plugging this into the $x$-component of the geodesic equation for the metric 
\eqref{eq:generalR} with the impulsive profile \eqref{eq:ip1} (derived under 
the same ``rules'' as above, see also \eqref{eq:geoRe} below), i.e.,
\begin{equation}
  \ddot x(u)+\frac{2k\theta(u)}{1+ku_+}\,\dot x(u)
\end{equation}
one does \emph{not} obtain a vanishing right hand side, but instead we have
\begin{equation} \label{eq:12} 
\ddot x(u)+\frac{2k\theta(u)}{1+ku_+}\,\dot x(u)=-\frac{kx_0\delta(u)}{(1+ku_+)^2} =-kx_0\delta(u),
\end{equation}
where in the last step we have used that for any function continuous 
in a neighbourhood of $u=0$ we have $f(u)\delta(u)=f(0)\delta(u)$. 

The result \eqref{eq:12} may be explained in rough terms also as follows: Since 
the $x$-component of the geodesics \eqref{eq:5.10} has a finite jump in its 
velocity at $u=0$ its second derivative will contain a term proportional to 
$\delta(u)$. On the other hand such a term cannot be present on the right hand 
side of  the geodesic equation in (R)-coordinates. Indeed the metric is 
Lipschitz continuous and hence possesses an $L^\infty$-connection: the 
Christoffel symbols in the geodesic equation will be proportional to the step 
function $\theta(u)$. Also the velocity $\dot x$ will only involve a step 
function, so that the $\delta$-term arising from $\ddot x $ cannot be cancelled.
\medskip

Third, one can resort to regularisation, which leads to different 
(and correct) geodesics, see item (A2) below, which again confirms that the 
geodesics of \eqref{eq:5.10} are not the right ones.
\medskip

But where is the error in the arguments of \cite[Sec.\ 5.1]{ZDH:impulsive}? 
It is included in the statement \eqref{eq:argument}. In fact, it simply makes no 
sense to relate the form of the geodesic equations in the before- and the 
after-zones in (S) coordinates. These are actually given on two non-overlapping 
patches and there is no way of relating quantities on either side but using the 
transformation to (R)-coordinates or alternatively to (B)-coordinates 
\emph{but} in this case keeping the distributional terms. 
\medskip

In conclusion, the geodesics \eqref{eq:5.10} cannot be seen as actually being 
geodesics of the spacetime \eqref{eq:generalR} with the impulsive profile 
\eqref{eq:ip1}. They do not satisfy the geodesic equations in any meaningful way 
across the impulse, neither in the sense of an appropriate solution concept, 
nor via a regualrisation approach, nor formally. Finally the physical 
argument \eqref{eq:argument} put forward in 
deriving \eqref{eq:5.10} is flawed.   

\subsection{Correctly deriving the geodesics using (R) \& (S)-coordinates}
Here we specialize the so-called $C^1$-matching procedure of \cite[Sec.\ 
3]{LSS:14} to the case of plane waves. This is a method to derive  ``jump 
conditions'' for the geodesics in impulsive waves as seen with respect 
to ``background coordinates'' in the before- and after-zones. We suspect that 
this was also the idea underlying the (flawed) approach of \cite[Sec.\ 
5.2]{ZDH:impulsive}.

In the particular case at hand we have a Minkowskian background in the before- 
and the after-zone and hence we can trivially derive the geodesics there in 
manifestly Minkowskian coordinates, i.e., in the (S)-coordinates\footnote{This 
  actually suggests a negative answer to question (Q1) above.}. We 
denote these geodesics in the usual $\pm$-notation as
\begin{equation}\label{eq:geo-s}
  \hat\xx^\pm(u),\quad \mbox{and}\quad \hat v^\pm(u).
\end{equation}
They are clearly just straight lines and are entirely determined by the following 
set of $2\times 6$ constants
\begin{align}\label{eq:const}
  \hat\xx^\pm_i:=\lim_{u\to 0\pm}\hat\xx^\pm(u),\quad 
  &\hat v^\pm_i:=\lim_{u\to 0\pm}\hat v^\pm(u)\nonumber\\
  \dot{\hat\xx}^\pm_i:=\lim_{u\to 0\pm}\dot{\hat\xx}^\pm(u),\quad 
  &\dot{\hat v}^\pm_i:=\lim_{u\to 0\pm}\dot{\hat v}^\pm(u),
\end{align}
where the subscript $i$ stands for ``interaction time'', i.e., for the instance when the 
geodesics cross the impulse. We can now relate the $\pm$-versions of 
these constants to one another using the fact that the geodesics in 
(R)-coordinates are 
$C^1$-curves. More precisely, we transform the geodesics \eqref{eq:geo-s} to 
(R)-coordinates in which we will denote them by 
\begin{equation}\label{eq:geoR}
 \xx(u),\quad \mbox{and}\quad v(u)
\end{equation}
and ``match'' the respective constants \eqref{eq:const}. To do so most 
explicitly we invoke the inverse transformation of \eqref{eq:trsfdiscont} to 
relate the (S)- to the (R)-coordinates\footnote{Note that we hence have to 
identify $\tilde x^\mu=(\tilde\xx,\tilde u,\tilde v)$ with $\hat 
x^\mu=(\hat\xx,\hat u,\hat v)$ in \eqref{eq:trsfdiscont}.}, which reads
\begin{align}\label{eq:invtrsf}
  \xx(u)=\PP^{-1}(u)\hat\xx(u),\quad
  v(u)=\hat v(u)+\frac{k}{2}\theta(u)\PP^{-1}(u)\hat\xx\cdot\J\hat\xx,
\end{align}
where again $\PP(u)=(\1+u_+k\J)$.
  
Now we can relate the $\pm$-versions of the constants \eqref{eq:const} as follows
\begin{align}\label{eq:junctionx}
  \hat\xx^-_i=\lim_{u\to 0-}\hat\xx^-(u)&=\lim_{u\to 0-}\xx(u)=
  \lim_{u\to 0+}\xx(u)\nonumber \\
  &=\lim_{u\to 0+}\PP^{-1}(u)\hat\xx^+(u)=
  \lim_{u\to 0+}(\PP^{-1}(u))\,\hat\xx^+_i=\hat\xx^+_i.
\end{align}
Here we have used the definition of $\hat\xx^-_i$ in the first equality, the 
transformation \eqref{eq:invtrsf} for $u<0$ in the second, the continuity of the 
geodesics \eqref{eq:geoR} in the third, and then again the transformation 
\eqref{eq:invtrsf}, now for $u>0$. Finally, we have used the definition of 
$\hat\xx^+_i$ and the explicit form of $\PP^{-1}$ to calculate the limit.

Similarly we may use the $C^1$-property of the geodesics \eqref{eq:geoR} 
to relate the respective \emph{velocities} on either side of the impulse and we 
obtain the following set of ``jump conditions'':
\begin{align}\label{eq:jump}
 \hat\xx^-_i&=\,\hat\xx^+_i,\qquad 
 &\hat v^-_i&=\,\hat v^+_i+\frac{k}{2}\hat\xx^+_i\cdot\J\hat\xx^+_i,
 \nonumber\\
 \dot{\hat\xx}^-_i&=\,\dot{\hat\xx}^+_i-k\J\hat\xx^+_i,\quad
 &\dot{\hat v}^-_i&=\,\dot{\hat v}^+_i+k\hat\xx^+_i\cdot\J\dot{\hat\xx}^+_-
 -\frac{k^2}{2}\hat\xx^+_i\cdot\J\hat\xx^+_i.
\end{align}
Observe that these relations are just a special case of the relations derived in \cite[Sec.\ 3]{LSS:14} with the same identifications as explained below \eqref{eq:trsfsplit}.
\medskip

We conclude with a remark on the ``philosophy'' of the $C^1$-matching, cf.\  
\cite[Rem.\ 4.1]{PSSS:15}. The matching \emph{presupposes} the following 
knowledge of the geodesics on the \emph{entire} spacetime: the geodesics heading
towards the impulse have to \emph{cross} it, have to be \emph{unique} and of \emph{$C^1$-regularity}. All these properties have been established for the situation at hand in \cite{LSS:14}. Also the $C^1$-matching procedure has been generalised to the case of non-expanding impulsive waves in any constant curvature background in \cite{PSSS:15} and to expanding impulsive waves, again in all constant curvature backgrounds in \cite{PSSS:16}.

\subsection{The geodesics in (B)-coordinates} Here we very briefly comment on 
the derivation of the geodesics in impulsive waves in (B)-coordinates. Indeed in 
\cite[Sec.\ 5.2]{ZDH:impulsive} the $\XX$-components of the geodesics in 
impulsive plane waves are derived by basically integrating the geodesic 
equations and the use of the ``multiplication rules'' \eqref{eq:mr} to yield
\begin{equation}\tag{5.16}\label{eq:5.16}
 \XX(U)=P(U)\XX_0=(\1+u_+k\J)\XX_0,
\end{equation}
where $\XX_0= \XX(0)$. Observe that \eqref{eq:5.16} is in perfect agreement with 
the left equations in \eqref{eq:jump}. The authors of \cite{ZDH:impulsive} 
correctly remark in footnote 11 on p.\ 12 that the derivation of the 
$V$-component is more involved from the distribution theoretic point of view.  

However, an ad-hoc procedure has been employed in the \emph{pp}-wave case to 
derive the geodesics in \cite{FPV:88}, which---to the author's best 
knowledge---is the earliest account explicitly calculating the geodesics in the 
distributional form of impulsive gravitational waves. A more reliable account 
has been put forward in \cite{Bal:97}, again in the \emph{pp}-wave case. Here 
some nonlinear theory of distributions was applied but still an ad-hoc 
assumption (preservation of the geodesic's tangent across the impulse) was 
needed to derive the result. The full solution was finally given in 
\cite{Ste:98a,KS:99a}. We remark that in these approaches, which essentially are 
based on regularisation of the impulsive profile by a sequence of general  
sandwich waves, it becomes a nontrivial task to show that the solutions of the 
now nonlinear geodesic equations live long enough to cross the regularised (and 
hence extended) wave zone, i.e., the impulse \emph{at all}. This is done using a 
fixed point argument which has been subsequently refined to allow a 
generalisation of the procedure to ever wider classes of impulsive waves, cf.\ 
\cite{SS:12,SS:15,SSS:16,SSLP:16,SS:17}.

\subsection{Returning to (R)-coordinates}\label{sec:3.5} In this section we 
finally comment on \cite[Sec.\ 7]{ZDH:impulsive}, where the geodesics 
\eqref{eq:5.7} are compared to the ones we have given in \cite[Sec.\ 
4]{Ste:98}\footnote{We remark that this note was prepared for the  
``Proceedings of the 8-th National Romanian Conference on GRG, Bistritza, 
June 1998''. However, it was never peer-reviewed and to   
the best of my knowledge the said volume never appeared, cf.\ the comment on 
ArXiv.}. 

In fact \cite{Ste:98} uses quite different conventions and there is a lapse in 
the geodesics presented in eq.\ (14) there---in fact the $X$- the $Y$- 
components should be interchanged. However, these geodesics have been correctly 
transferred to the present setting in \cite[Sec.\ 7]{ZDH:impulsive} to read 
(using (R)-coordinates $\xx=(x,y)$) \begin{equation}\tag{7.1}\label{eq:7.1}
 x(u)=x_0+\dot x_0\left(\frac{u_+}{1+u_+}+u_-\right),\quad
 y(u)=y_0+\dot y_0\left(\frac{u_+}{1-u_+}+u_-\right),\quad 
\end{equation}
where for brevity we again restrict attention to the spatial components leaving 
aside the more complicated $v$-equation. Also we set $k=1$ and use the usual 
definition $u_-=u$ if $u\leq 0$ and $u_-=0$ for $u\geq 0$.

Actually we are aware of three ways to directly derive the explicit form of the 
geodesics in (R)-form in the plane wave case, i.e. \eqref{eq:7.1}, all without 
the use of the symmetries of the spacetime and all leading to the same result:
\begin{enumerate}
 \item[(A1)] Solving the geodesic equations, which are given e.g.\ in \cite[eq.\ (7.3)]{ZDH:impulsive} and explicitly read
 \begin{equation}\label{eq:geoRe}
  \ddot x+2\frac{k\theta}{1+u_+}\dot x=0,\quad
  \ddot y-2\frac{k\theta}{1-u_+}\dot y=0,
 \end{equation}
separately in the before- and the after-zones and matching the integration 
constant to obtain a global $C^1$-curve. \item[(A2)] Regularising the step 
function in the metric e.g.\ by setting 
$\theta_\eps(u)=\int_{-\infty}^u\rho_\eps(t)dt$ (with 
$\rho_\eps\to\delta$ in distributions),  
then integrating the regularised equations, and finally performing the limit 
$\eps\to 0$.
\item[(A3)] Making an ad-hoc ansatz (essentially guessing the solutions) and 
checking that the equations do hold again using the ``multiplication rules'' 
\eqref{eq:mr}.
\end{enumerate}

In \cite[Sec.\ 4]{Ste:98} we actually only mention approaches (A1) and (A2). 
However, the fact that the $C^1$-property is used in the matching is explicitly 
stated above eq.\ (14)---contrary to the claim made below \cite[eq.\ 
(7.1)]{ZDH:impulsive}. Anyhow, it has meanwhile been proven that the 
$C^1$-property holds, cf.\ Section \ref{sec:3.3}, above.    

Moreover, approach (A2) and hence also indirectly the $C^1$-property is 
confirmed in \cite[Caption of Fig.\ 7]{ZDH:impulsive}, where the authors 
acknowledge the fact that a regularisation by Gaussians leads to geodesics 
converging to \eqref{eq:7.1}. 
\medskip

Finally, we extend the calculations of eqs.\ \eqref{eq:10}--\eqref{eq:12} by 
formally showing that also the geodesics \eqref{eq:5.7} (with arbitrary initial 
speeds) do \emph{not} satisfy the geodesic equations \cite[eq.\ 
(7.3)]{ZDH:impulsive}, i.e., \eqref{eq:geoRe}. Indeed the spatial components of 
\eqref{eq:5.7} take the explicit form \begin{equation}\tag{7.2}\label{eq:7.2}
 x(u)=x_0+\dot x_0^+\frac{u_+}{1+u_+}+\dot x_0^-u_-,\quad
 y(u)=y_0+\dot y_0^+\frac{u_+}{1-u_+}+\dot y^-_0 u_-, 
\end{equation}
and using again the usual set of ``multiplication rules'' \eqref{eq:mr} one obtains e.g.\ for the $x$-component in $-\infty<u<1$ that
\begin{equation}
 \ddot x+2\frac{k\theta}{1+u_+}\dot x=\delta(u)\,(\dot x^+_0-\dot x^-_0).
\end{equation}
This equation again tells us that in order to satisfy the geodesic equation we need to have $\bigtriangleup\dot\xx=\dot\xx^+_0-\dot\xx^-_0=0$, i.e., no jumps in the velocities of the geodesics in (R)-coordinates.

\subsection{Comparing the geodesics in (B)- and (R)-coordinates}\label{sec:3.7}
In the final section of this chapter we comment on \cite[Sec.\ 
5.3]{ZDH:impulsive} and the interrelations between the geodesics in (R)-form 
and in (B)-form. 
The authors of \cite{ZDH:impulsive} say on this matter:
\medskip

\begin{quote}
{\it 
``The naive expectation might be that this [interrelation] could be achieved by 
using the  transformation formula between the coordinates, (2.6), i.e., 
\begin{equation}\tag{5.20}
\XX(U)=\PP(U)\,\xx(u),
\label{naiveXx}
\end{equation}
which is indeed correct in the case of continuous wave profiles for particles initially at rest, \cite{ZDGH:pl,ZDGH:soft}, for which $\xx(u)=\xx_0=\mbox{const}$ 
for all $u$. However, identifying the initial positions,
$ 
\xx_0=\XX_0
$ 
and  combining (\ref{eq:5.16}) and (\ref{eq:5.10}) yields instead, 
\begin{equation}\tag{5.21}
\medbox{
\XX(U)= (\PP^T\!\PP)(u)\, \xx(u) = a(u)\, \xx(u)\,.\,
}
\label{goodXx}
\end{equation}
Where does the extra $\PP$-factor come from? The clue is that 
\emph{the delta-function $\delta(u)$  makes the velocity jump}  both in B [(B)] and BJR [(R)] coordinates  --- and does it in the \emph{opposite} way, see in (5.15)\footnote{Equation (5.15) in \cite{ZDH:impulsive} reads $\dot\XX(0+)=c_0\XX_0$.} and (5.12b) [that actually should read (5.12a)], respectively. 
The extra $\PP$ factor takes precisely care of these jumps:
 the first $\PP$ in (\ref{goodXx}) straightens the trajectory (5.10) to the trivial one, 
$
\PP(u)\xx(u)=\xx_0,
$
which has zero initial BJR velocity  as in the smooth case \cite{ZDGH:pl,ZDGH:soft};
 then the second $\PP(u)$ factor curls it up according to (\ref{naiveXx}), 
yielding $\XX(u)$ in (5.16).''
}
\end{quote}
\medskip

This explanation remains dubious. While it is true that the combination of (\ref{eq:5.16}) and (\ref{eq:5.10}) yields \eqref{goodXx}, this just confirms that the geodesics in (R)-form \eqref{eq:5.10} are \emph{not} the correct ones. In fact, replacing the incorrect geodesics \eqref{eq:5.10} by the correct ones, i.e., \eqref{eq:7.1},  which in the case at hand, i.e., vanishing speeds, simply read $\xx(u)=\xx_0$, we correctly obtain \eqref{naiveXx} (as is also acknowledged in the above quotation):
\begin{equation}
 \XX(U)=\PP(U)\XX_0=\PP(u)\xx_0=\PP(u)\xx(u).
\end{equation}

The fact that formally transforming the geodesics in (R)-form \eqref{eq:7.1} 
with the ``discontinuous change of coordinates'' \eqref{eq:trsfdiscont} 
yields exactly the geodesics in (B)-form \eqref{eq:5.16} has already been noted 
in \cite[Sec.\ 4]{Ste:98} just below eq.\ (14)\footnote{Note that in the last 
line on p.\ 7 there is a typo: the equation number (11) should actually be 
(10).}.  In fact, it \emph{does nothing else but} transforming the (B)-geodesics 
with vanishing initial speeds into the (R)-geodesics. In other words, the broken 
and jumping (B)-geodesics become the new coordinate lines in the (R)-system. And 
this is the ultimate reason why the regularity of the metric improves from 
distributional in the (B)-coordinates to  continuous in (R)-coordinates. 
\medskip

The formal calculation establishing these ideas has been turned into a solid 
piece of mathematics even for the \emph{pp}-wave case in \cite{KS:99}, using 
nonlinear distributional geometry. A good way to describe the situation in 
physical terms is given there in Sec.\ 5: The ``discontinuous change'' of 
coordinates is the distributional limit of a family of smooth transformations 
which can be obtained by a general regularisation procedure, which is adapted to 
the spacetime geometry.  From this regularisation point of view, the (B) and 
(R)-forms of the impulsive metric arise as the distributional limits of the 
\emph{same} sandwich wave in different coordinate systems. In such a scenario, 
in general, different spacetimes may result and the fact that in this case the 
geometries are ``physically equivalent'' is reflected by the fact that the 
resulting transformation is merely discontinuous rather than unbounded. 
Nevertheless, it introduces finite jumps of the geodesics and their velocities. 
\bigskip

\section{Summary and Conclusions}

We have clarified the intricacies of the particle motion in impulsive 
plane---and effectively in \emph{pp}--waves. In (B)-coordinates the 
geodesics possess a discontinuous $v$-component and the $v$-velocity as well as 
the transverse velocities exhibit a finite jump across the impulse. Then again 
in (R)-coordinates the geodesics are continuously differentiable curves and 
hence there is \emph{no} jump in the velocities. This seemingly odd behaviour 
is due to the fact that the transformation between the (B)- and 
(R)-coordinates is discontinuous. It nevertheless allows one to correctly 
and consistently transform the geodesics (formally) from one form to the other. 
Moreover, this procedure has been handled in a mathematically meaningful way 
using nonlinear distributional geometry. 

All this is in perfect agreement with the geometric picture: The 
(R)-coordinates are \emph{comoving} and hence discontinuous as seen from the 
two Minkowski halves to either ``side'' of the impulse. This is why the 
regularity of the metric improves from distributional in (B)-coordinates to 
continuous in (R)-coordinates. But it is also the reason why one does not see 
any particle motion in (R)-coordinates: Particles initially at rest remain so 
after the impulse until they eventually reach the coordinate singularity of the 
(R)-coordinates. 
\medskip

Finally, this author remains agnostic regarding the question whether or not 
these results mean that impulsive plane waves exhibit a ``velocity memory 
effect''. The reason simply is that  we are not aware of an invariant 
definition of the ``memory effect'' for the spacetimes at hand, which are 
\emph{not} asymptotically flat.

\section*{acknowledgements} I wish to thank my frequent coauthors Ji\v{r}\'{i}  
 Podolsk\'y, Robert \v{S}varc, and Clemens S\"amann  for their 
encouragement during my writing of this note and for their ongoing support, and 
Michael Kunzinger for the critical reading of the manuscript. Also 
I acknowledge a very friendly email conversation with Peter Horvathy. This work 
was supported by project P28770 of the Austrian Science Fund FWF and the 
WTZ-grant CZ12/2018 of OeAD.

\bibliographystyle{abbrv}
\bibliography{ro}

\begin{thebibliography}{10}

\bibitem{Bal:97}
H.~Balasin.
\newblock Geodesics for impulsive gravitational waves and the multiplication of
  distributions.
\newblock {\em Classical Quantum Gravity}, 14(2):455--462, 1997.

\bibitem{BH:03}
C.~Barrab\`es and P.~A. Hogan.
\newblock {\em Singular null hypersurfaces in general relativity}.
\newblock World Scientific Publishing Co., Inc., River Edge, NJ, 2003.
\newblock Light-like signals from violent astrophysical events.

\bibitem{BG}
V.~B. Braginsky and L.~P. Grishchuk.
\newblock Kinematic resonance and the memory effect in free mass gravitational
  antennas.
\newblock {\em Sov. Phys. JETP}, 62:427, 1985.

\bibitem{BT}
V.~B. Braginsky and K.~S. Thorne.
\newblock Present status of gravitational-wave experiments.
\newblock In {\em Proceedings of the ninth international conference on general
  relativity and gravitation ({J}ena, 1980)}, pages 239--253. Cambridge Univ.
  Press, Cambridge, 1983.

\bibitem{C}
D.~Christodoulou.
\newblock Nonlinear nature of gravitation and gravitational-wave experiments.
\newblock {\em Phys. Rev. Lett.}, 67(12):1486--1489, 1991.

\bibitem{CG:12}
P.~T. Chru{\'s}ciel and J.~D.~E. Grant.
\newblock On {L}orentzian causality with continuous metrics.
\newblock {\em Classical Quantum Gravity}, 29(14):145001, 32, 2012.

\bibitem{DGHZ:caroll}
C.~Duval, G.~W. Gibbons, P.~A. Horvathy, and P.-M. Zhang.
\newblock {Carroll symmetry of plane gravitational waves}.
\newblock {\em Class. Quant. Grav.}, 34(17):175003, 2017.

\bibitem{FPV:88}
V.~Ferrari, P.~Pendenza, and G.~Veneziano.
\newblock Beam-like gravitational waves and their geodesics.
\newblock {\em Gen. Relativity Gravitation}, 20(11):1185--1191, 1988.

\bibitem{Fil:88}
A.~F. Filippov.
\newblock {\em Differential equations with discontinuous righthand sides},
  volume~18 of {\em Mathematics and its Applications (Soviet Series)}.
\newblock Kluwer Academic Publishers Group, Dordrecht, 1988.
\newblock Translated from the Russian.

\bibitem{GKSS:19}
J.~D.~E. Grant, M.~Kunzinger, C.~S\"amann, and R.~Steinbauer.
\newblock The future is not always open.
\newblock {\em preprint, arXiv:1901.07996 [math.DG]}, 2019.

\bibitem{GP:09}
J.~B. Griffiths and J.~Podolsk\'y.
\newblock {\em {Exact Space-Times in Einstein's General Relativity}}.
\newblock Cambridge University Press, Cambridge, 2009.

\bibitem{GKOS:01}
M.~Grosser, M.~Kunzinger, M.~Oberguggenberger, and R.~Steinbauer.
\newblock {\em Geometric theory of generalized functions with applications to
  general relativity}, volume 537 of {\em Mathematics and its Applications}.
\newblock Kluwer Academic Publishers, Dordrecht, 2001.

\bibitem{KS:99}
M.~Kunzinger and R.~Steinbauer.
\newblock {A note on the Penrose junction conditions}.
\newblock {\em Class. Quant. Grav.}, 16:1255--1264, 1999.

\bibitem{KS:99a}
M.~Kunzinger and R.~Steinbauer.
\newblock A rigorous solution concept for geodesic and geodesic deviation
  equations in impulsive gravitational waves.
\newblock {\em J. Math. Phys.}, 40(3):1479--1489, 1999.

\bibitem{LSS:14}
A.~Lecke, R.~Steinbauer, and R.~\'{S}varc.
\newblock {The regularity of geodesics in impulsive pp-waves}.
\newblock {\em Gen. Rel. Grav.}, 46:1648, 2014.

\bibitem{P:68a}
R.~Penrose.
\newblock Structure of space-time.
\newblock In {\em Battelle Rencontres, 1967 Lectures in Mathematics and
  Physics}, pages 121--235. Benjamin, New York, 1968.

\bibitem{P:68}
R.~Penrose.
\newblock {Twistor quantization and curved space-time}.
\newblock {\em Int. J. Theor. Phys.}, 1:61--99, 1968.

\bibitem{P:72}
R.~Penrose.
\newblock The geometry of impulsive gravitational waves.
\newblock In {\em General relativity (papers in honour of {J}. {L}. {S}ynge)},
  pages 101--115. Clarendon Press, Oxford, 1972.

\bibitem{P:02}
J.~Podolsk\'y.
\newblock {Exact impulsive gravitational waves in space-times of constant
  curvature}.
\newblock In {\em Gravitation: Following the Prague Inspiration}, pages
  205--246. Singapore: World Scientific Publishing Co., 2002.

\bibitem{PSSS:15}
J.~Podolsk{\'y}, C.~S{\"a}mann, R.~Steinbauer, and R.~{\v{S}}varc.
\newblock The global existence, uniqueness and {${C}^1$}-regularity of
  geodesics in nonexpanding impulsive gravitational waves.
\newblock {\em Classical Quantum Gravity}, 32(2):025003, 23, 2015.

\bibitem{PSSS:16}
J.~Podolsk{\'y}, C.~S{\"a}mann, R.~Steinbauer, and R.~{\v{S}}varc.
\newblock The global uniqueness and {$C^1$}-regularity of geodesics in
  expanding impulsive gravitational waves.
\newblock {\em Classical Quantum Gravity}, 33(19):195010, 23, 2016.

\bibitem{SS:12}
C.~S\"amann and R.~Steinbauer.
\newblock On the completeness of impulsive gravitational wave spacetimes.
\newblock {\em Classical Quantum Gravity}, 29(24):245011, 11, 2012.

\bibitem{SS:15}
C.~S\"amann and R.~Steinbauer.
\newblock Geodesic completeness of generalized space-times.
\newblock In {\em Pseudo-differential operators and generalized functions},
  volume 245 of {\em Oper. Theory Adv. Appl.}, pages 243--253.
  Birkh\"auser/Springer, Cham, 2015.

\bibitem{SS:17}
C.~S\"amann and R.~Steinbauer.
\newblock Geodesics in nonexpanding impulsive gravitational waves with
  {$\Lambda$}. {II}.
\newblock {\em J. Math. Phys.}, 58(11):112503, 18, 2017.

\bibitem{SSLP:16}
C.~S\"amann, R.~Steinbauer, A.~Lecke, and J.~Podolsk\'y.
\newblock Geodesics in nonexpanding impulsive gravitational waves with
  {$\Lambda$}, part {I}.
\newblock {\em Classical Quantum Gravity}, 33(11):115002, 33, 2016.

\bibitem{SSS:16}
C.~S\"amann, R.~Steinbauer, and R.~\v{S}varc.
\newblock Completeness of general {$pp$}-wave spacetimes and their impulsive
  limit.
\newblock {\em Classical Quantum Gravity}, 33(21):215006, 27, 2016.

\bibitem{Sou:73}
J.-M. Souriau.
\newblock Le milieu \'elastique soumis aux ondes gravitationnelles.
\newblock {\em {C}olloq. {I}nternat.\ {CNRS}}, 220:243--256, 1974.
\newblock Avec discussion.

\bibitem{Ste:98a}
R.~Steinbauer.
\newblock Geodesics and geodesic deviation for impulsive gravitational waves.
\newblock {\em J. Math. Phys.}, 39(4):2201--2212, 1998.

\bibitem{Ste:98}
R.~Steinbauer.
\newblock On the geometry of impulsive gravitational waves.
\newblock {\em ArXiv:9809054[gr-qc]}, 1998.

\bibitem{Ste:14}
R.~Steinbauer.
\newblock Every {L}ipschitz metric has {$C^1$}-geodesics.
\newblock {\em Classical Quantum Gravity}, 31(5):057001, 3, 2014.

\bibitem{W:98}
W.~Walter.
\newblock {\em Ordinary differential equations}, volume 182 of {\em Graduate
  Texts in Mathematics}.
\newblock Springer-Verlag, New York, 1998.
\newblock Translated from the sixth German (1996) edition by Russell Thompson,
  Readings in Mathematics.

\bibitem{ZP}
Y.~B. Zel'dovic and A.~G. Polnarev.
\newblock {Radiation of gravitational waves by a cluster of superdense stars}.
\newblock {\em Sov.\ Astron.}, 18:17, 1974.

\bibitem{ZDGH:soft}
P.-M. Zhang, C.~Duval, G.~W. Gibbons, and P.~A. Horvathy.
\newblock {Soft gravitons and the memory effect for plane gravitational waves}.
\newblock {\em Phys. Rev.}, D96(6):064013, 2017.

\bibitem{ZDGH:pl}
P.-M. Zhang, C.~Duval, G.~W. Gibbons, and P.~A. Horvathy.
\newblock {The Memory Effect for Plane Gravitational Waves}.
\newblock {\em Phys. Lett.}, B772:743--746, 2017.

\bibitem{ZDH:impulsive}
P.-M. Zhang, C.~Duval, and P.~A. Horvathy.
\newblock Memory effect for impulsive gravitational waves.
\newblock {\em Classical Quantum Gravity}, 35(6):065011, 20, 2018.

\bibitem{ZEGH:18}
P.-M. Zhang, M.~Elbistan, G.~Gibbons, and P.~A. Horvathy.
\newblock Sturm-{L}iouville and {C}arroll: at the heart of the memory effect.
\newblock {\em Gen. Relativity Gravitation}, 50(9):Art. 107, 9, 2018.

\end{thebibliography}

\end{document}